# Radial analysis and scaling of housing prices in French urban areas

Gaëtan LAZIOU[a], Rémi LEMOY[a] and Marion LE TEXIER[b]

[a]University of Rouen, IDEES Laboratory UMR 6266 CNRS, Mont-Saint-Aignan, France;

[b]University Paul Valery Montpellier 3, LAGAM, Montpellier, France.

e-mail: gaetan.laziou1@univ-rouen.fr, remi.lemoy@univ-rouen.fr, marion.le-texier@univ-montp3.fr

**Abstract**

Urban scaling laws summarize how urban attributes evolve with city size. Recent criticism questions notably the aggregate view of this approach, which leads to neglecting the internal structure of cities. This is all the more relevant for housing prices due to their important variations across space. Based on a dataset compiling millions of real estate transactions over the period 2017-2021, we investigate the regularities of the radial (center-periphery) profiles of housing prices across cities, with respect to their size. Results are threefold. First, they corroborate prior findings in the urban scaling literature stating that largest cities agglomerate higher housing prices. Second, we find that housing price radial profiles scale in three dimensions with the power 1/5 of city population. After rescaling, great regularities between radial profiles can be observed, although some locational amenities have a significant impact on prices. Third, it appears that our rescaled profiles approach fails to explain housing price variations in the city center across cities. In fact, prices near the city center rise much faster with city size than those in the periphery. This has strong implications for low-income households seeking homeownership, because prohibitive prices in the center may contribute to pushing them out into peripheral locations.

## Introduction

Today, more than half of the world's population lives in urban settlements, and urbanization is still an ongoing trend (UNPD, 2018). Thus, it is of growing interest to understand how cities evolve to eventually make them more sustainable. One of the concerns relates to the emergent "urban housing affordability crisis", as depicted by some researchers (Wetzstein, 2017). This calls for a greater understanding and consciousness of affordability challenges in the real estate market, as well as their social and geographical implications (Le Goix et al., 2021).

Housing affordability depends on the value of residential properties, and then on their characteristics. Since Rosen (1974), an extensive body of literature has discussed the implicit price of several characteristics on home value. Among this literature, many authors assess the impact of distance to the Central Business District (CBD) on housing prices, as the bid-rent theory from Alonso (1964) predicts a decreasing trend from the center (see the review by Yiu and Tam, 2004). Even if some authors found that the distance to the CBD is not a significant driver of housing prices after controlling for structural and neighborhood characteristics (e.g. Waddell et al., 1993), other studies suggest that the housing price gradient corresponds well to the expectations derived from the Alonso model, as in Stavanger and Berlin (Ahlfeldt, 2011; Osland et al., 2007).

Unfortunately, analyses carried out on a large sample of cities remain limited, due to lack of data. We can however highlight that some authors have pursued this direction. For instance, Kulish et al. (2012) offer a view on housing prices with respect to the distance to the CBD for the five largest cities in Australia. More recently, Liotta et al. (2022) found that some predictions of the monocentric framework are verified for a global sample of cities of more than 300,000 inhabitants. Other studies show that specific patterns (or, at least, regularities) in housing prices can be found across cities. In Italy, d'Acci (2023) recently observed a scale invariance in the housing price distribution across cities: the housing price of a municipality decays in a predictive way with its rank within a region. Many studies, whether in geography or economics, also focus on how prices vary between urban areas (e.g. Potepan, 1996; Sarkar, 2019). In France, a comprehensive research is proposed by Combes et al. (2019) who estimated the elasticity of urban costs at the center of cities with respect to population. Here, we contribute to this body of literature, by examining the radial variations of housing prices (radial profiles) across cities of different sizes (scaling laws).

To investigate how an urban indicator evolves through cities of difference sizes, scaling laws have benefited from a renewed interest these last years, since the work of Pumain (2004) and Bettencourt et al. (2007). A growing number of studies look at cities as complex systems whose shape follows specific scaling laws (Batty, 2008). These are expressed as power-law relationships connecting an urban attribute and city size (usually measured by its population). The scaling exponent summarizes the variation of an urban attribute with city size: it can be either constant across cities (linear regime), more concentrated in the upper-part of the urban hierarchy (superlinear regime), or conversely in the lower-part (sublinear regime).

Yet, recent criticism toward this approach has been raised. One point relates to the aggregation of quantities at city level. This is all the more relevant as urban geography lacks a robust definition of what a city is. Based on a comprehensive literature review, Batty & Ferguson (2011) state that scaling laws are subject to uncertainties, because definition of cities have changed over time, and call for a new definition of the city. More recently, Louf & Barthelemy (2014) and Cottineau et al. (2017) further show that urban scaling is relative to the definition of cities, as distinct delineations can lead to opposite conclusions.

Besides, studying aggregate quantities overlooks their distribution within cities. For instance, Sarkar (2019) and Shutters et al. (2022) examine the scaling of wages and housing costs changes through categories. This approach considers intra-urban statistical disparities, but still lacks a view on the geographical variations within cities. In the urban scaling literature, only few authors investigate the internal structures of cities, for instance through their radial profiles, in both theoretical and empirical analyses (Delloye et al., 2020; Lemoy & Caruso, 2020).

In this paper, we conduct an empirical analysis to understand how the distribution of housing prices changes as city population grows, and also investigate the scaling of radial profiles. We rely on the monocentric tradition from well-known models in urban geography and economics (Alonso, 1964 ; Clark, 1951; Fujita, 1989; Von Thünen, 1826), and consider cities through concentric rings. Such an analysis is made possible by the increasing availability of large-scale data on housing prices. While some researchers work on geolocated data collection from real estate websites for intra-urban analyses (e.g. Chapelle and Eyméoud, 2022), we use here an online and open-access dataset with millions of real estate transactions.

The remainder of this article is organized as follows: first, we present the geographical scope of the analysis, our dataset and method; second, we detail our results on scaling laws and investigate the robustness of generic profiles as well as non-radial structures; finally, we discuss the implications of our findings, and some perspectives of this work.

## Data and methods

*Geographical scope*

Following the standards of the above urban scaling literature, we consider a functional definition of cities. Specifically, we use the latest delineations of urban areas provided by the French National Institute of Statistics and Economic Studies (INSEE). We then compute the 2020 population $N$ of urban areas based on municipal-level census data.

The top left panel of Figure 1 shows the (counter-cumulative) distribution of city sizes in France $P(x \geq N)$. Not surprisingly, a large proportion of cities has a small population, while only a few cities have a high population. The distribution is characterized by a power-law tail, showing a slight deviation from Zipf's law for cities. On the contrary, the distribution of cities above 50,000 inhabitants follows a power-law with an estimated exponent of -1.01, which is consistent with Zipf's rank-size rule.

We can further observe the relationship between the number of sales in our dataset and population from the top-right panel of Figure 1. These two variables are almost proportional (scaling exponent close to 1), which indicates that the real estate market is not necessary more dynamic in large cities. Given the volume of transactions per city over the period 2017-2021, we focus our radial and scaling analysis on those with a population exceeding 50,000 inhabitants.

This sample of cities above 50,000 inhabitants provides a wide diversity of situations, in terms of city size, attractiveness, location in France (Figure 1, bottom). From this sample, we remove cities for which all or part of the data is not available (departments of Bas-Rhin, Haut-Rhin and Moselle, located in the East of France), as well as urban areas whose center is located abroad (e.g., Geneva, Monaco), and overseas France. We end up with 161 urban areas, representing 49.8 million inhabitants, around three quarters of France's population.

Note that we distinguish, for the radial analysis, 'ordinary' cities from some other cities. The latter present characteristics likely to impact prices, including coastal cities, mountainous cities, defined as cities whose center is located in Europe's longest mountain range zone (Alps), cities along the Swiss border and polycentric cities[1]. They account for 75 cities in total.

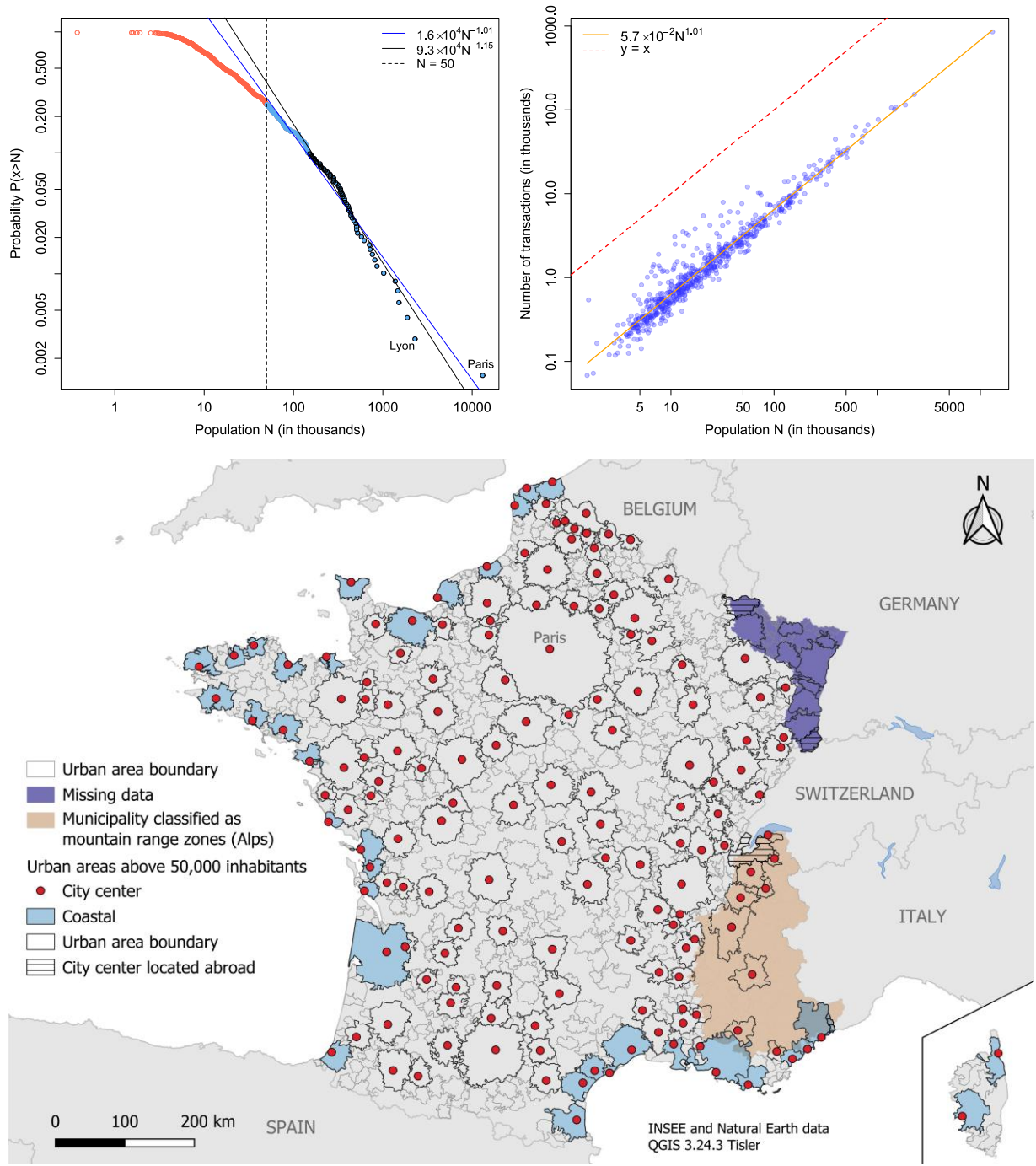

**Figure 1**. Top left panel: counter cumulative distribution of city sizes in the database $P(x \geq N)$ using population counts from census data. Top right panel: number of transactions recorded in the geolocated DVF dataset from January 2017 to December 2021 as a function of population. Bottom panel: Map of urban areas in France.

## Presentation and preparation of the DVF dataset

Our main material is the *Demande de Valeurs Foncières* (DVF) dataset, produced by the French Central State tax services. It provides information on land and real estate sales over the past five years, and is considered as a reference for property transaction in France. The main assets of the dataset lie in its exhaustiveness and nationwide coverage. Note that the dataset does not rely on asked price, unlike now widely-used ads scrapped from real estate websites (e.g. Chapelle and Eyméoud, 2022). Some information is available on the transaction itself (date of transfer, nature of the transfer, net selling price, etc.), on the precise location (address, land registry reference, etc.), and on structural characteristics (living area, number of main rooms, type of property).

In this paper, nationwide data from January 2017 to December 2021 is used (variations across time of the number of sales are displayed in supplementary material Figure *S1*). Overall, the dataset contains 7,037,983 transactions (number of unique identifiers). However, we remove transactions including commercial or industrial premises (which do not follow the same market principles). Second, we choose not to keep sales involving two or more apartments or houses, neither those with three or more outbuildings. To account for the upward trend in housing prices over the period, we reassess the price according to the housing price index provided by the INSEE, taking the first quarter of 2020 as a reference (see supplementary material Figure *S2*). Finally, we compute the price per square meter (m²) and remove peculiar transactions[2] as well as obvious outliers (price per square meter under 250 €/m² or above 50,000 €/m²). We end up with 4,297,231 transactions nationwide (see supplementary material Table *S1* for descriptive statistics of the dataset).

### Radial and scaling analysis

There is a positive skewness in housing price distribution at national scale (see supplementary material Figure *S3*). This may suggest wide variations between small-, medium-sized and large cities. We first seek to observe how the distribution changes with respect to population. For this purpose, we conduct a scaling analysis of housing price categories on all urban areas through a power function of population size, following the equation:

$$Y = \alpha N^\beta, \tag{1}$$

where $\alpha$ is a normalization constant (theoretical value of $Y$ for $N = 1$) and $\beta$ a scaling exponent under study, reflecting the general rule across the system of cities. For convenience, the relationship is regressed in its log-transformed form $\log(Y) = \log(\alpha) + \beta \log(N)$. Following the recommendations of Leitao et al. (2016), we use the Bayesian Information Criterion (BIC) to check a statistical evidence for $\beta \neq 1$. More precisely, we compare the maximum-likelihood $\mathcal{L}$ of each model to the corresponding model with a fixed $\beta = 1$. The difference between the two models can be written as follows:

$$\Delta \text{BIC} \equiv \text{BIC}_{\beta=1} - \text{BIC}_\beta. \tag{2}$$

If $\Delta \text{BIC} < 0$, then the model with fixed $\beta$ is better, while if $0 < \Delta \text{BIC} < 6$ the results are inconclusive, and if $\Delta \text{BIC} > 6$ the model with $\beta \neq 1$ is better.

However, the originality of our methodology lies in investigating the internal heterogeneity of housing prices within urban areas by performing a radial and scaling analysis on a sample of 161 cities. We express a three-dimensional scaling law for the radial function of housing prices $p(r)$ with the homothetic form:

$$p(r) = N^\beta g\left(\frac{r}{N^\beta}\right), \tag{3}$$

where $N$ is the population of each city, $\beta$ a scaling exponent and $r$ the Euclidean distance to the center. Note that this equation corresponds to a three-dimensional scaling, with a rescaling on the two (horizontal) Euclidean dimensions of geographical space (distance to the center $r = (x^2 + y^2)^{1/2}$) and on the vertical dimension of price.

In detail, we extract the transactions within the boundaries of the urban areas. Then, we define concentric rings of fixed width 1km around the city center and then aggregate the price per square meter of transactions within each ring at distance $r$ to the city center. Similarly to Kulish et al. (2012), we compute the median housing price, under the requirement that more than ten transactions occurred within a ring.

As the concentric rings are defined from the city center, its location is an important choice. In the monocentric model of Alonso (1964), Mills (1967) and Muth (1969), the city center is the CBD, i.e. the place where jobs are located and residents commute to. In the literature, various methodologies have been used to locate the CBD. In this work, we choose the historical city hall as the city center, because it has proven to be a robust choice to shed light on the internal structure of cities, for land uses and population densities (Broitman and Koomen, 2020; Lemoy and Caruso, 2020), but also for prices (Atack and Margo, 1998).

Once radial profiles are computed, we rescale them. As seen from equation (3), we use the total population $N$ as a scaling parameter, because population statistics have proved to be robust for establishing scaling effects (Pumain, 2004). The value of the exponent $\beta$ thus provides an estimate of the population-elasticity of housing price in the city center (vertically) and along the profile (horizontally).

## Results: scaling laws

In this section, we investigate housing prices between ('macro level') and within ('meso level') cities with respect to population. For this purpose, we sequentially perform a scaling analysis of the housing price distribution and study the scaling of radial profiles of housing prices.

### Housing price distribution

We perform a scaling analysis of housing price categories in all French urban areas through an ordinary least squares (OLS) approach. Note that we define ten categories (classes of 250€/m² and 500€/m²) with similar transaction volumes, following the distribution of price per square meter.

On Figure 2, results show that while lower housing price categories scale sublinearly or linearly with city size, we face a superlinear regime when it comes to high housing price categories. This statement remains true even if we take into account the 95% confidence interval and the $\Delta$BIC following the urban scaling literature. In other words, most affordable cities are mainly concentrated at the bottom of the urban hierarchy (i.e. small-sized cities), while it is the opposite for high housing prices.

These results support the claim that large cities are much less affordable for households, especially the least affluent ones. They corroborate the findings of Sarkar (2019) in Australian SUAs and USA MSAs, as we are lead to the same conclusion that "housing costs rise disproportionately with city size". However, since the scaling behavior of wages and wealth creation is also known to be superlinear (Bettencourt et al., 2007; Shutters et al., 2022), it might be inferred that high housing costs are offset by higher wages. Nevertheless, homeownership for low-income households living in largest cities is made difficult, as they compete with wealthier households in the real estate market.

Note that fluctuations may be considered as important, as error bars are large. But they can be partly explained by a few locational amenities, primarily the coastal one (see supplementary material Figure *S4* and *S5* for scatterplots with best fits and maps of residuals). Setting outliers aside, this scaling analysis shows that housing prices across the system of cities follow specific patterns. However, we lack a geographical perspective on the intra-urban structure of housing prices. Here we aim to fill this gap.

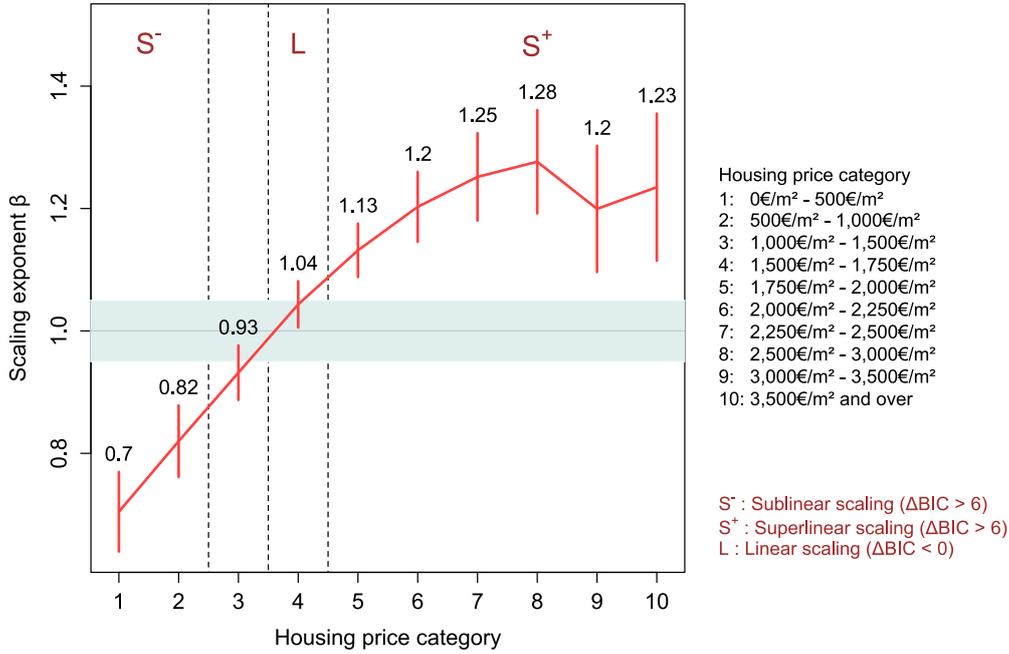

**Figure 2**. Scaling analysis of housing price categories for French urban areas. Each point indicates the exponent $\beta$ of a linear regression of the logarithmic number of transactions in a given housing price category with respect to logarithmic population (the value of $\beta$ is written in black). Error bars represent the 95% confidence interval of the exponent, and the colored area represents the interval [0.95; 1.05].

## Housing price radial profiles

We denote $p(r)$ as the median housing price at distance $r$ from the city center. A salient observation from the left column of Figure 3 (where the radial profiles of 'ordinary' cities are displayed) is the sharp decrease tendency of housing prices from the city center to the periphery, especially in large cities. It is in line with expectations of the monocentric tradition and the residential choice-based theory of Alonso, which states that housing prices result from households' trade-off between transportation costs to employment centers (i.e. the CBD) and housing consumption. Because curves appear clustered by population size, a scaling effect is visible: the biggest city (Paris, top of urban hierarchy, with more than 13 million inhabitants) appears at the top, followed by two regional metropolitan areas with more than 1 million inhabitants, Lyon and Toulouse, two large cities (Rennes and Angers), two medium-sized cities, located in rural areas (Poitiers and Auxerre) and two small cities (Castres, Fougères). This suggests that city population is a relevant variable to explain price disparities between cities, whether in the city center or in the periphery.

It is thus interesting to rescale the horizontal and vertical axes proportionally to a power of the city population. This is done on the right column of Figure 3. The most homogeneous profiles between cities are obtained with a rescaling parameter which is the power 1/5 of city population, $N^{1/5}$. Then, the rescaled distance to the center is $r' = r \times (N_{Paris}/N)^{1/5}$ while the rescaled housing price is $p'(r') = p(r') \times (N_{Paris}/N)^{1/5}$. This corresponds to $\beta$ = 1/5 in equation (3). We note that this value $\beta$ = 1/5 is very close to the elasticity of 0.208 identified by Combes et al. (2019) with other methods and data. The legend of the top right panel indicates the rescaling parameter $k = (N_{Paris}/N)^{1/5}$. Note that, because we use Paris as a reference, $k_{Paris} = 1$, while $k_{Angers} = 2$ ($N = 4.38 \times 10^5$) and $k_{Fougères} = 3$ ($N = 5.20 \times 10^4$). The rescaling seems to work satisfactorily with the sample of cities within $r'$ ranging from 10 to 60 km from the city center as it captures a clear common trend.

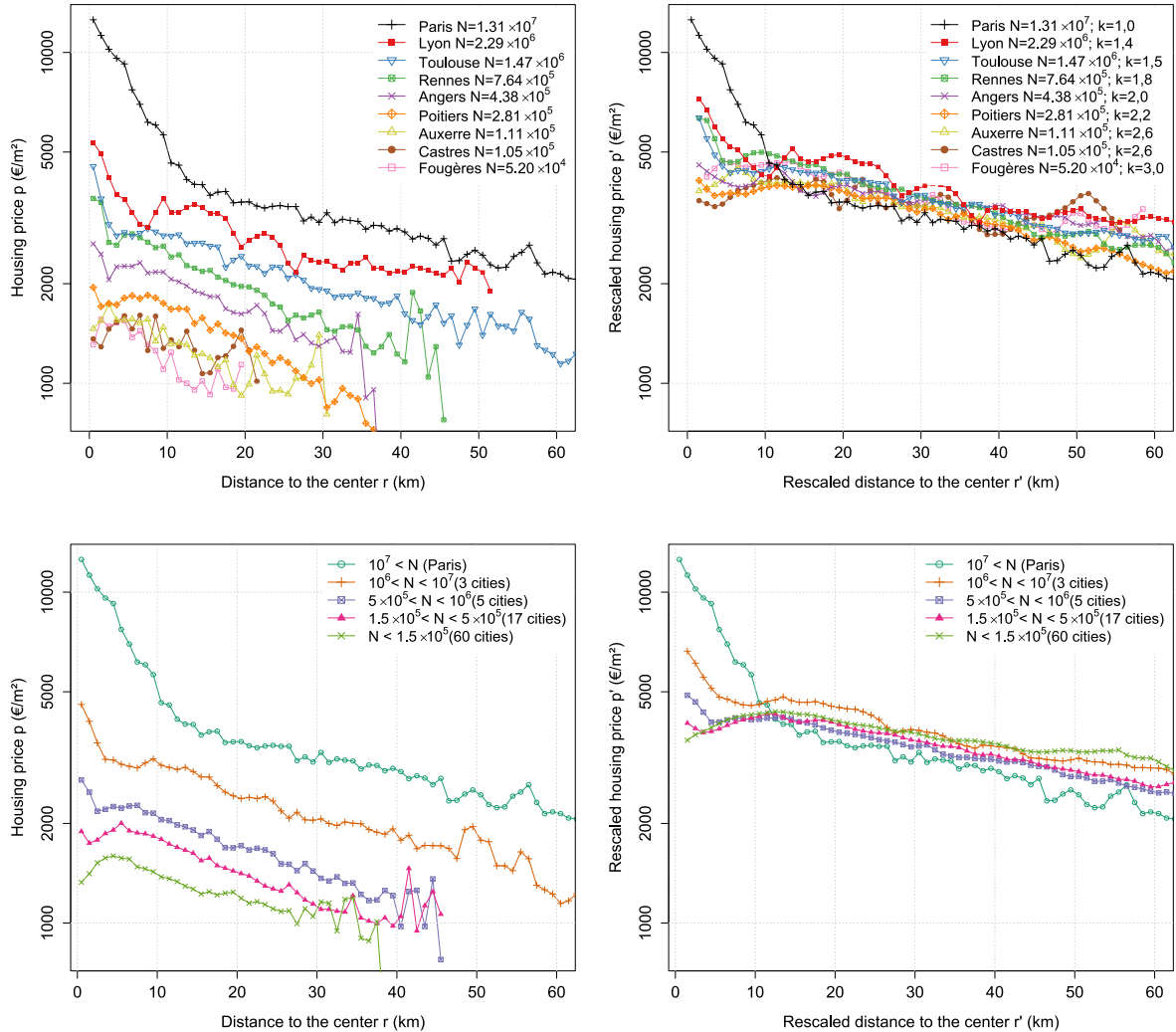

**Figure 3**. Left column: Housing price as a function of the distance to the center in a selection of cities of different sizes (top) and in 'ordinary' cities in different population categories (bottom). The 2020 population $N$ is given in the legend. Right column: rescaled curves for the same cities – on the top panel, the rescaling factor $k$ (s.t. $r' = rk$ and $p' = pk$) is given in the legend.

Yet, near the city center curves are still clustered by population, which is obvious on the bottom right panel of Figure 3, where all 'ordinary' cities are displayed by city size. Thus, the rescaling captures most of the radial profile but fails near the city center, which calls for further investigation. To address this point, we examine the power law relationship connecting housing prices in the city center and population. For this purpose, we look at housing price $p_c$ within a distance of $0 < r < 1$ km from the center. However, population is very unevenly distributed in the system of cities. Concerning this point, Leitao et al. (2016) indicate that "when data are viewed in the usual double logarithmic plot, the best curve will be the one that passes close to most points, i.e. it weights a village as much as a million-size city". We show in supplementary material Figure *S7* that the classic log-log regression performs unsatisfactorily, because it fails to predict housing price in large cities. To account for this limitation, we also perform Weighted Least Square (WLS) regressions, using as weights population $N$ ('person model') and the square of population $N^2$.

The scaling relationship performed on all 161 cities between median housing price at the center $p_c$ and city population $N$ is displayed on Figure 4. The classic log-log regression gives a scaling exponent of $\beta = 0.29 \pm 0.03$ (95% CI, R² = 0.38). Yet, this is an artifact of the presence of cities with specific characteristics, because the fit provides a poor prediction for 'ordinary' cities. Instead, we trust the weighted log-log regressions of housing price in the center, $p_c$, against population $N$, which cancels the unequal distribution of city sizes. The best fits return much higher exponents, $\beta = 0.44 \pm 0.01$ (95% CI, R² = 0.88) and $\beta = 0.50 \pm 0.01$ (95% CI, R² = 0.94) using as weights population $N$ and square population $N^2$, respectively. We conclude that prices near the city center increase much more quickly with population than in the rest of the profile. Note also that these values are far from the elasticity of 0.208 identified by Combes et al. (2019) at the city center, with other methods and data.

Several conclusions can be drawn from the scaling of urban housing price profiles. Morphological patterns of housing prices show strong regularities across cities after rescaling, which indicates a rather clear homothety of housing price profiles in French cities. The exponent 1/5 of the city population which we find here can be compared to the exponent 1/3 observed by Lemoy & Caruso (2020) for population density in European cities: it is lower, but not much. Thus, prices along the radial profile do not increase as much as population density increases: large cities are less affordable than small ones, but one could have expected a higher exponent, in agreement with population density and thus demand size. On the contrary, near the city center the different (and much greater) exponent indicates that homeownership there becomes more difficult as the size of the city increases, much more than in the periphery.

What also emerges from the analysis is the unconventional behavior of cities with locational amenities (such as a coastal or mountainous areas), as illustrated by Figure 4. They are a clear source of fluctuation in this scaling relationship, as they very often exhibit higher housing prices than 'ordinary' cities, whether in the city center or in the periphery. These results are in agreement with a line of literature stating that, beyond city size, there are other drivers for prices, as the distance to the coastline, socio-demographic characteristics or income (e.g. Ahlfeldt, 2011; Ayouba et al., 2021). Thus, some small or medium towns with highly-valued amenities and services are also hardly affordable for the poorest.

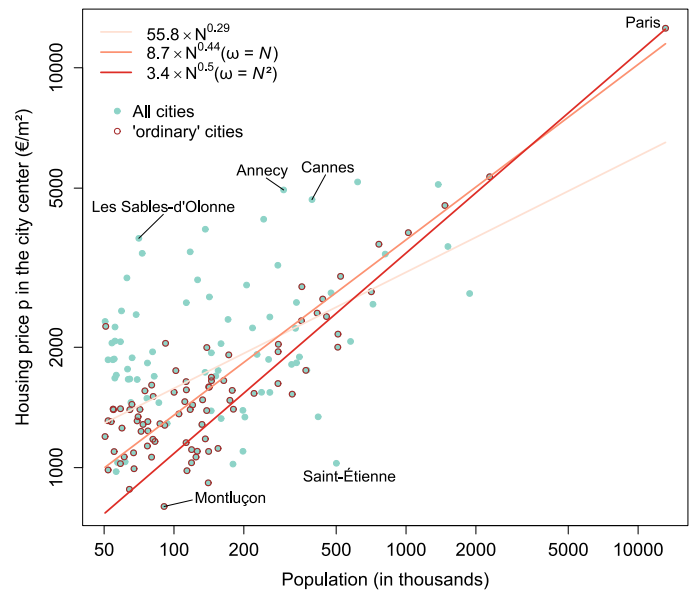

**Figure 4**. Median housing price in the center ($0 < r < 1$ km) of French cities as a function of city population. Regressions are performed on all cities. In the second and third regressions, cities are weighted according to their population $N$ and square population $N^2$, respectively.

*Optimal scaling exponent*

The rescaling captures a common trend, even though fluctuations are important (see Supplementary material Figure *S9*). To check if 0.2 is the optimal scaling exponent, we assess the accuracy (and performance) of the scaling of radial profiles using two different indicators. We test a wide range of exponents $\beta$ for both the rescaled distance to the center $r' = r \times (N_{Paris}/N)^\beta$ and the housing price $p'(r') = p(r') \times (N_{Paris}/N)^\beta$. Note that in order to have data at an identical rescaled distance $r'$ for every city, we perform a simple linear interpolation at 1km interval.

We first measure the effectiveness of the scaling with the signal over noise ratio (SNR), already used by Lemoy & Caruso (2020). It measures the relative dispersion along the radial profile and follows the equation:

$$SNR = \frac{\sum_{r'=0}^{r'=r_f} p'(r')}{\sum_{r'=0}^{r'=r_f} \sigma\big(p'(r')\big)},$$ (4)

with $p'(r')$ the signal, defined by the average housing price after rescaling, and $\sigma(p'(r'))$ the noise, i.e. the standard deviation of housing prices after rescaling. As the maximum rescaled distance differs between cities, some profiles are missing in the periphery (see supplementary material Figure *S10*): we denote as $r_f$ the distance at which $\frac{n+1}{2}$ curves have ended. The rescaling is successful if the signal is important while the noise is low, so we look for a high SNR value.

Besides, we use a complementary indicator that measures the clustering or ranking of curves with population. For this purpose, we create a new indicator, namely the Root Mean Square Correlation (RMSC), which aims at measuring whether the curves are ordered according to city size or not, after rescaling. We compute at each rescaled distance to the center $r'$ and after interpolation, the correlation coefficient between the logarithm of cities' total population and the rescaled housing price $p'$. If the absolute correlation is low, it means that the rescaling is successful in taking off the size effect. The interpolation method and the number of points on which the indicator is calculated remain unchanged. Thus, the RMSC follows the equation:

$$RMSC = \sqrt{\frac{\sum_{r'=0}^{r'=r_f} R(p'; \log(N))^2}{n}}$$ (5)

with $p'$ the rescaled housing price, $N$ the population, $n$ the number of data points, and $r_f$ as defined above. Given that the rescaling intends to remove the city size effect, we expect the RMSC to be close to zero, unlike the SNR.

Figure 5 shows that the exponent yielding the highest SNR and the lowest RMSC is close to 0.2, i.e. the exponent used in our analysis. In detail, we observe that the mean price along the profile is more than three times higher than the standard deviation after rescaling, if we consider all cities of the database. Not surprisingly, the SNR is much higher if we consider 'ordinary' cities only, as outliers are taken off. On the other hand, the RMSC is especially low for $\beta$ = 0.2, indicating that curves are not ranked by city size (i.e. the population effect is removed). Conversely, without rescaling ($\beta$ = 0) the RMSC is high, due to the scaling law of housing price radial profiles.

We note that the optimal exponent is clearly 0.2 if we consider only cities above 150,000 inhabitants, but slightly lower for all cities. This can be explained by polycentricity and specific urban amenities, as they sometimes lead to high prices even in small cities (thus mitigating the optimal scaling exponent).

In the supplementary material figures *S11* to *S15*, we further show that the radial profiles of these peculiar cities may exhibit non-radial structures.

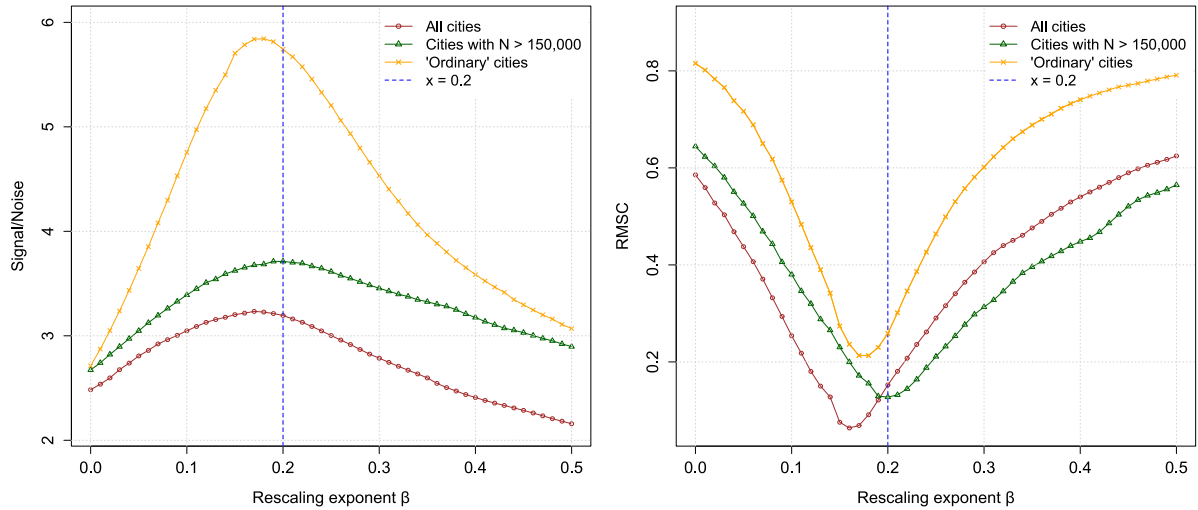

**Figure 5.** Signal over noise ratio (left panel) and Root Mean Square Correlation (right panel), for different values of scaling exponent $\beta$. The vertical dashed line indicates $\beta = 0.2$.

## Discussion

This study investigates the behavior of housing prices through the French system of cities, and hence contributes to the radial and urban scaling literature. The relevance of this analysis stems from the fact that housing prices have proved to vary considerably both within and across cities. Our results provide new insights on the homothety of housing prices across cities and along radial profiles.

We have established that while the number of transactions scales linearly with city size (highlighting a proportionality between market dynamics and population), highest housing prices clearly agglomerate in larger cities. Yet, we note that our combined center-periphery and scaling analysis provides much more information on the internal structure of cities than the traditional urban scaling analysis shown on Figure 2, by taking the intra-urban space into account. We find that the radial profiles of urban housing prices are homothetic (in the two Euclidean dimensions of geographical space and in the vertical dimension of price), and scale in three dimensions with the power 1/5 of city population.

The regularity of this scaling law of housing price profiles can be considered both expected and quite surprising, depending on the point of view. Expected because it is quite similar to what Lemoy & Caruso (2020) observe for land use and (especially) population density. Surprising because housing price is a very different variable, much less 'physical' and much more 'anthropic', since it is largely determined by a housing market. We could have expected much less regularity for housing prices, even when setting aside peculiar cities.

This homothetic scaling of price in three dimensions is quite close to the one observed for population density by Lemoy and Caruso (2020) for European urban areas, although the exponent is different: 1/5 for price versus 1/3 for population density. Taking as a reference the largest city of the dataset (i.e. Paris), this means that medium and small cities have a further-reaching influence on housing prices in their periphery than on population density. This will raise issues in polycentric areas, since the profiles of different cities can be expected to overlap rather easily. For this reason, France is a very interesting case study, because the country has very few conurbations. We note than $1/5 \approx 0.2$ and $1/3 \approx 0.33$ are rather close values of exponents, but still different, as can be seen on Figure 5. Another important difference is the fact that housing price variations between the center and the periphery are much smaller than population density variations: a factor of the order of 1.5 (small-sized city) to 5 (Paris) for prices versus 20 to 200 for population density.

Regarding the theory which could explain these stylized facts, the Alonso-LU model proposed by Delloye et al. (2020) might actually be rather convincing, although it predicts a minimal exponent of 1/3 for prices (instead of 1/5 observed here) and uses a 1/3 exponent for land use (instead of the observed 1/2) in order to obtain a perfect 3-dimensional scaling of population density. We note that housing price and population density could have a rather direct relationship since they have quite similar variations (see Lemoy et al., 2011). One reason for the difference between them, is that in the Alonso-LU model land use is incorporated in (raw) population density, which is why researchers sometimes study net population density, where natural land uses are taken off. And we know that radial profiles of artificial land use are exponential (Lemoy and Caruso, 2021). But natural land use is completely ignored in our study of (median) housing price: this variable has no such direct influence on housing price as it has on population density.

The regularities in housing prices' radial profiles which we observe here confirm the very strong role of total population $N$, which we use as a scaling parameter. Once this parameter is fixed, it seems that almost everything is fixed in terms of the urban center-periphery structure: land use, population density, housing price, transport (Mennicken et al., 2023). From the knowledge of just this total population $N$, a lot can be said about almost any city.

In the city center however, our results clearly indicate that prices are more sensitive to city population than the whole profile, as we find a higher exponent (0.4-0.5 or the center versus 0.2 for the whole profile). These figures highlight that housing prices in the city center increase sharply with city size, while the increase is slower in peripheral locations. This phenomenon is linked to the attractiveness of urban cores, which leads to a stronger competition between households for central places (Lemoy et al., 2010). Its evolution in time in particular needs to be further studied with other datasets. Even if the ongoing densification observed in some countries could allow more households to settle in the city center (Broitman et Koomen, 2020), our results suggest prohibitive prices in the city center of large cities. This could contribute to pushing out low-income households seeking homeownership into peripheral locations, and lead to socio-spatial segregation. In France, it has recently been shown that access to homeownership for the poorest has sometimes come at the cost of living far from city centers, leading to a "suburbanization of poverty" (Gobillon et al., 2022). Further research could be carried out in this direction.

More generally, this work raises quite many empirical and theoretical research questions regarding housing prices, which could be studied in further work, depending on data availability. How do these scaling laws translate to other geographical spaces (countries, continents)? How do they evolve in time? What about housing rents instead of purchasing prices? What about land prices? What is the reason of the specific behavior of city centers? Is this compatible with urban economic theory (Alonso model, Alonso-LU model)? How should conurbations be specifically considered in this analysis: just as independent cities close to each other, or with characteristic interactions between their housing markets and radial profiles? How could coastal cities be specifically modelled?

These scaling laws of housing price (plural because we actually observe two laws, one for the center and another for the rest of the profile) are rather bad news for planners trying to keep central areas of large cities accessible for all households, not just wealthy ones. They deliver a message of powerlessness of policymakers facing these phenomena, as we find great regularities of housing prices in the city centers across cities. We hope that it can be received as a warning that this issue of high housing prices in large cities' centers is a serious one, which should be dealt with using appropriate means.

## Notes

1. More precisely, we draw buffers around city centers with a radius of $80 \times (\frac{N}{N_{Paris}})^{1/3}$ kilometers.

   A city is considered polycentric if more than 20% of a city's buffer is covered by one or more other buffers.

2. We remove exchanges, expropriations, auction sales and sales of building land.

## Supplementary material

### 1. Volume of transactions

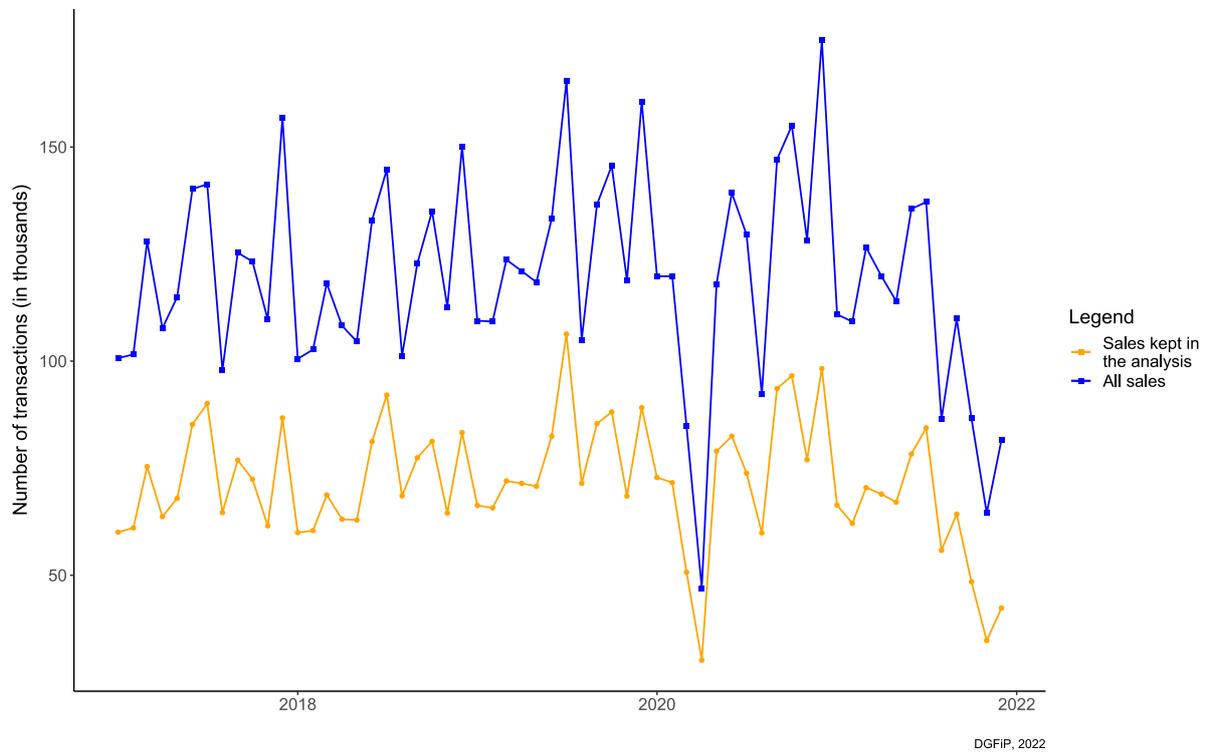

*Figure S1. Number of transactions over time in France (from January 2017 to December 2021).*

### 2. Descriptive statistics

The dataset we use for the radial and scaling analysis includes 4,297,231 transactions. The main characteristics of these transactions are detailed below.

| | Mean | Sd | Min | 25% | 50% | 75% | Max |
|---|---|---|---|---|---|---|---|
| Type (%): House | 55.6 | | | | | | |
| Type (%): Apartment | 44.4 | | | | | | |
| Nature (%): Sale | 96.2 | | | | | | |
| Nature (%): VEFA | 3.8 | | | | | | |
| Price per sqm | 2,950 | 2,382 | 250 | 1,495 | 2,318 | 3,592 | 49,863 |
| Surface | 82 | 42 | 9 | 54 | 76 | 101 | 1640 |
| Number of rooms | 3.5 | 1.5 | 0 | 2 | 3 | 4 | 112 |
| Number of outbuildings | 0.4 | 0.6 | 0 | 0 | 0 | 1 | 2 |

*Table S1. Descriptive statistics of the DVF dataset*

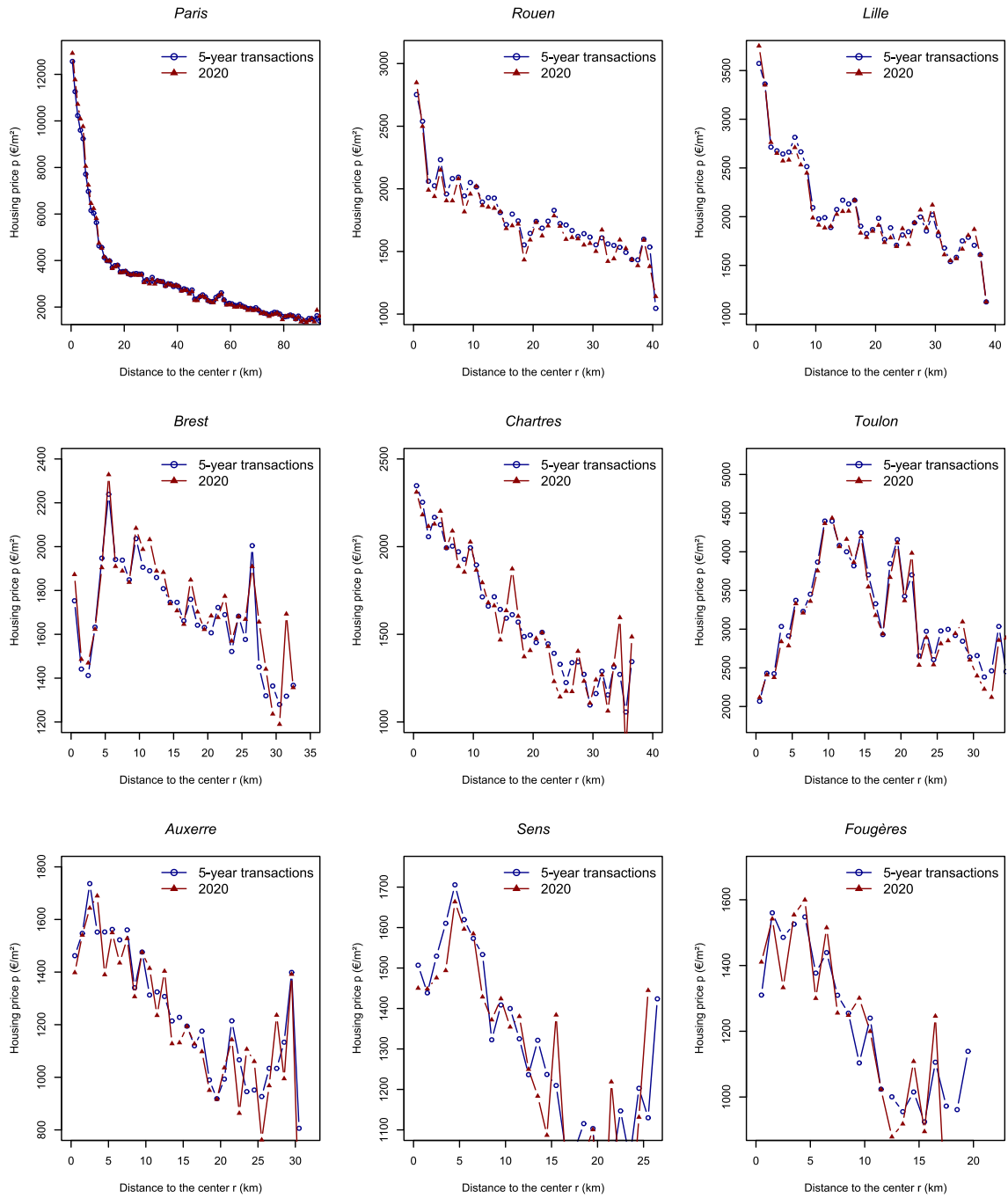

*Figure S2. Housing price as a function of the distance to the center in a sample of cities of different sizes. Radial profiles are displayed for transactions that took place in 2020 only, and transactions over the period 2017-2021*

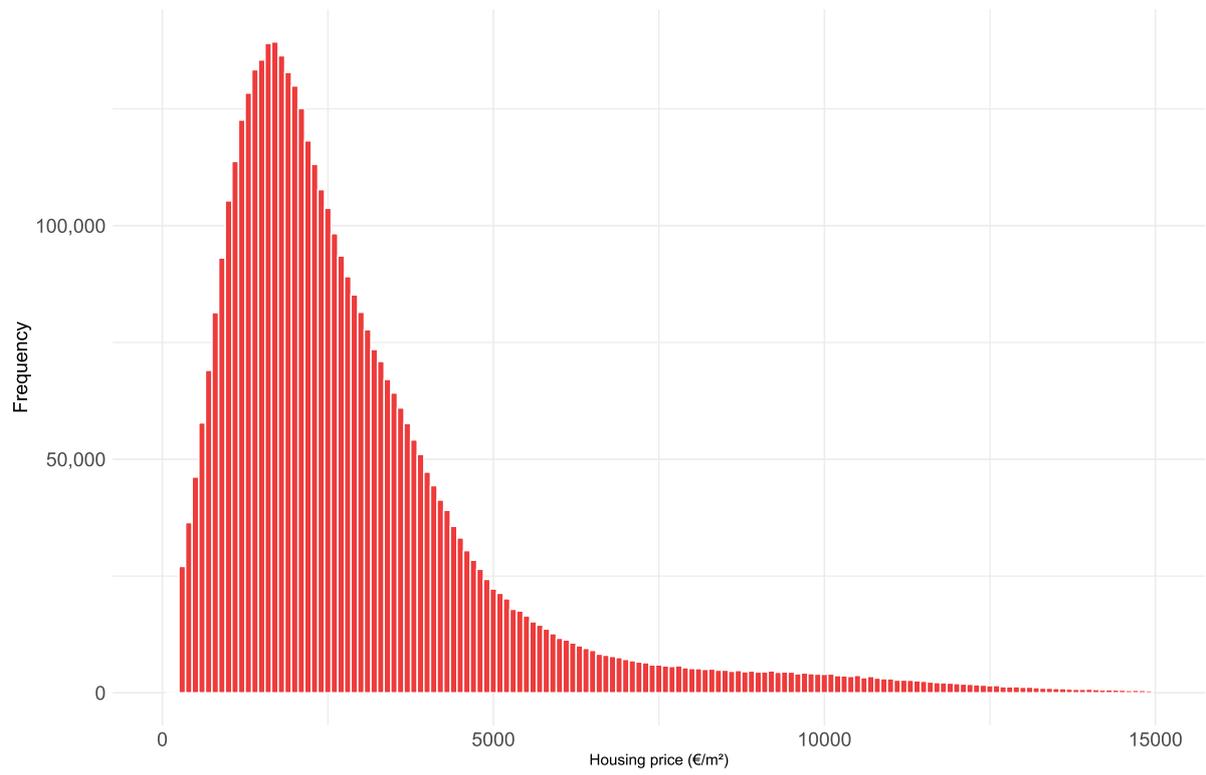

*Figure S3. Housing price distribution in France (€/m²)*

## 3. Scaling analysis of housing price categories

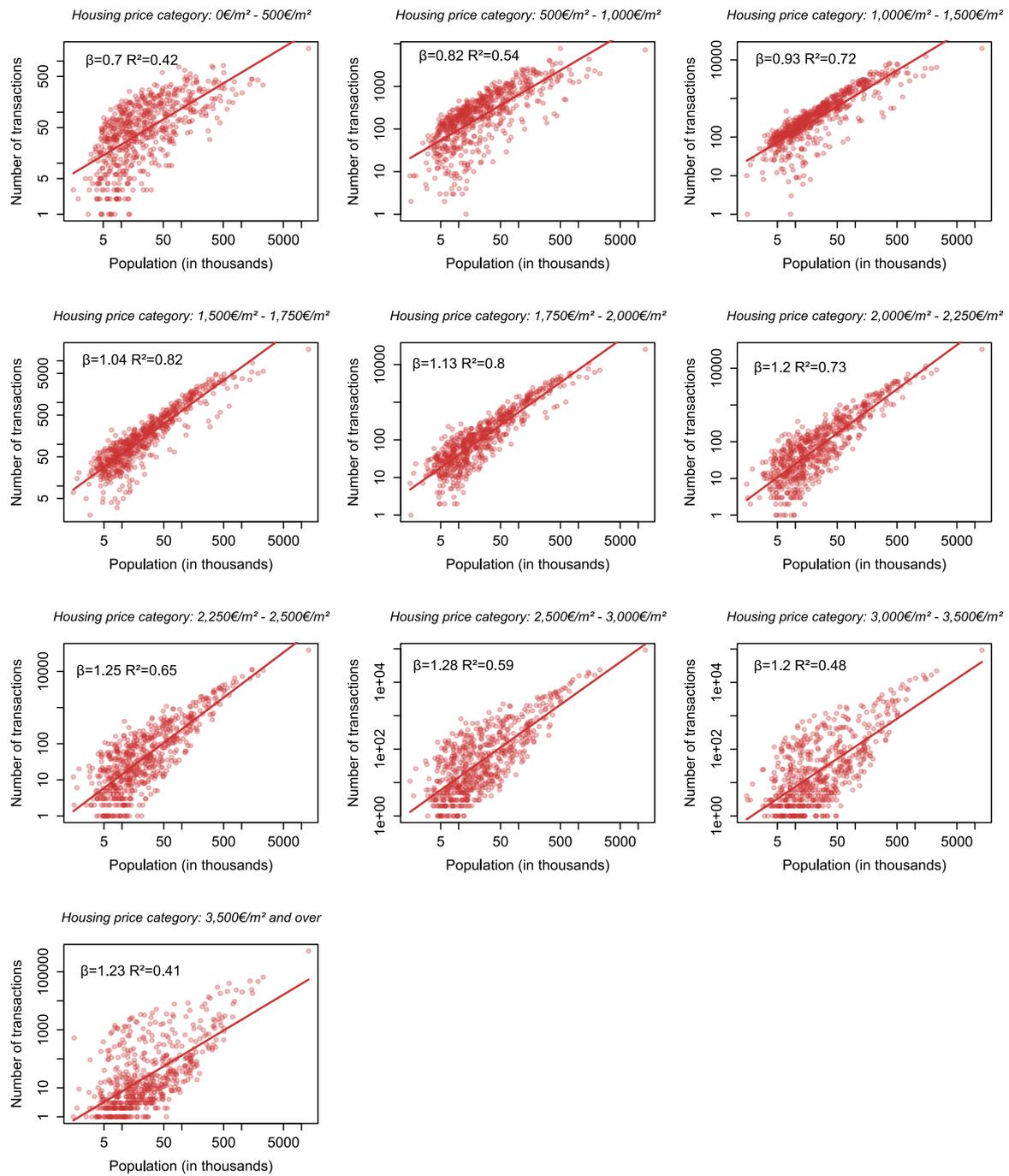

*Figure S4. Scaling analysis of housing price categories for French urban areas. Note that zero values are excluded from the analysis – however, there are very few of them.*

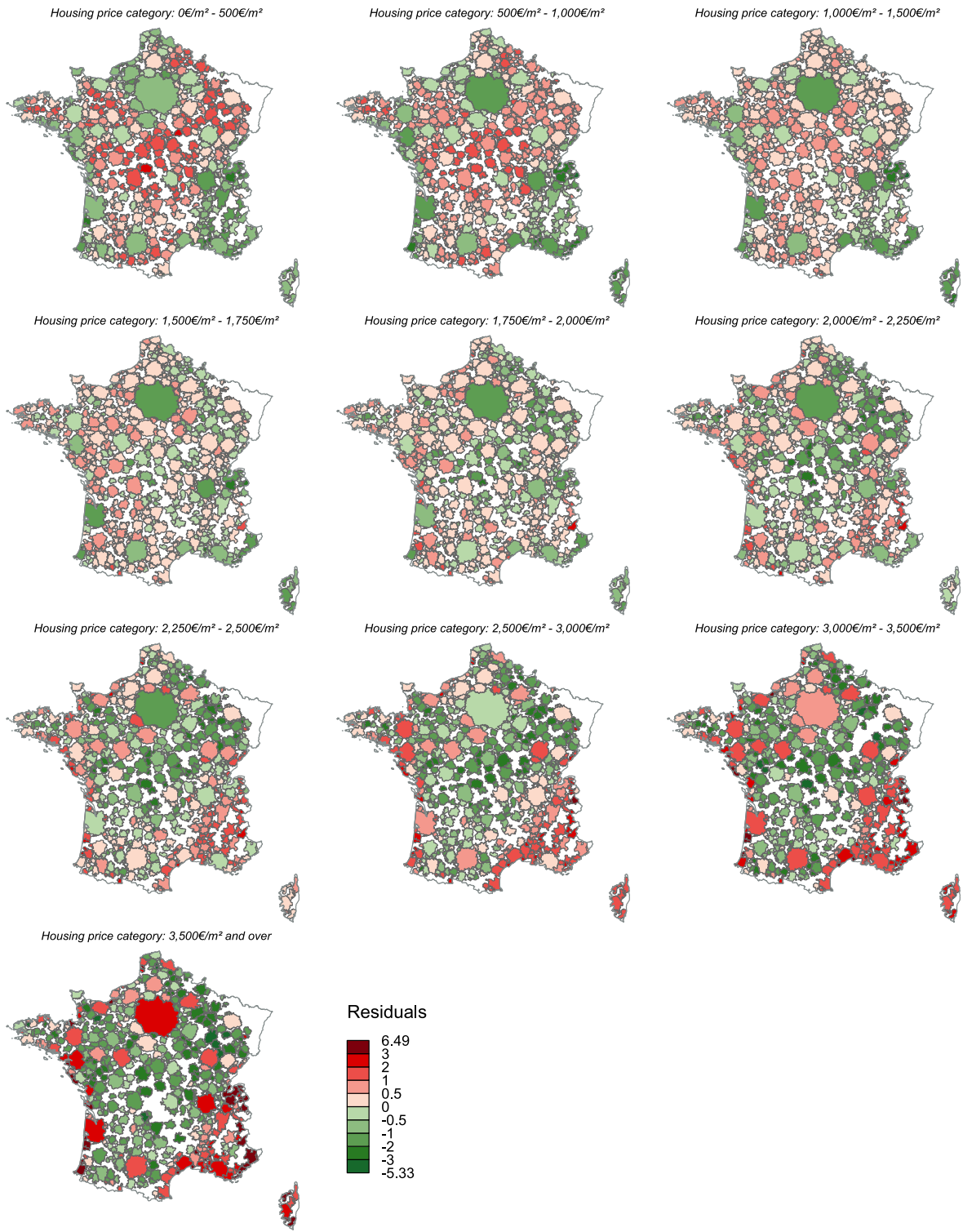

*Figure S5. Residuals from the scaling analysis of the logarithmic number of transactions in a given housing price category with respect to logarithmic population Y= $\alpha N^\beta$ (eq. 1 of the main text). Residuals are obtained by a regression in logs. Red values indicate a higher number of transactions in reality than expected with respect of population, and conversely for green values. Dark colors mean large deviations.*

## 4. Scaling of housing price radial profiles (houses only)

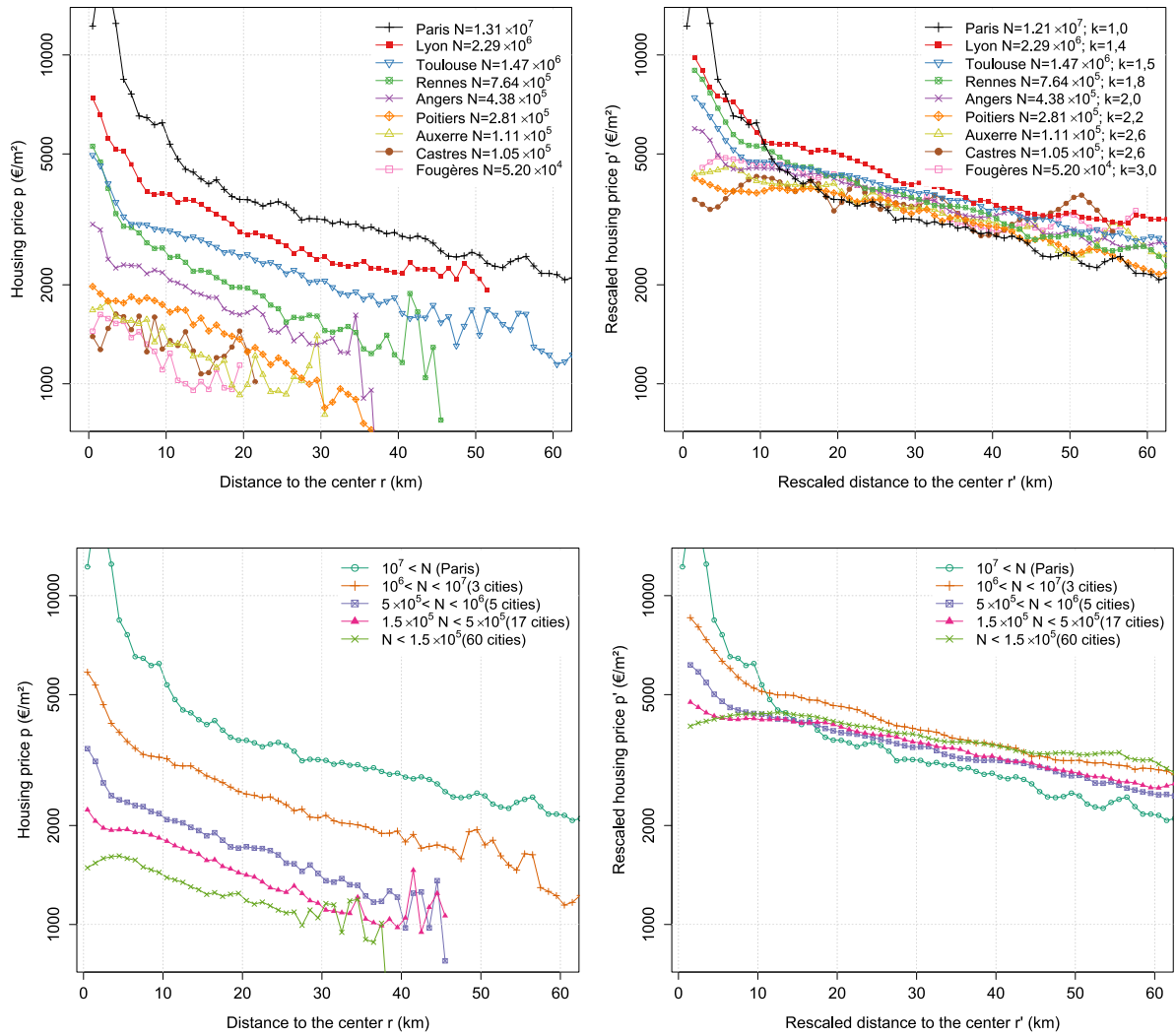

*Figure S6. Left column: Housing price as a function of the distance to the center in a selection of cities of different size (top) and in 'ordinary' cities of different population categories (bottom). The 2020 population N is given in the legend. Right column: rescaled curves for the same cities – on the top panel, the rescaling factor k (s.t. r' = rk and p' = pk) is given in the legend. Profiles are displayed for houses only.*

## 5. Weighted regressions

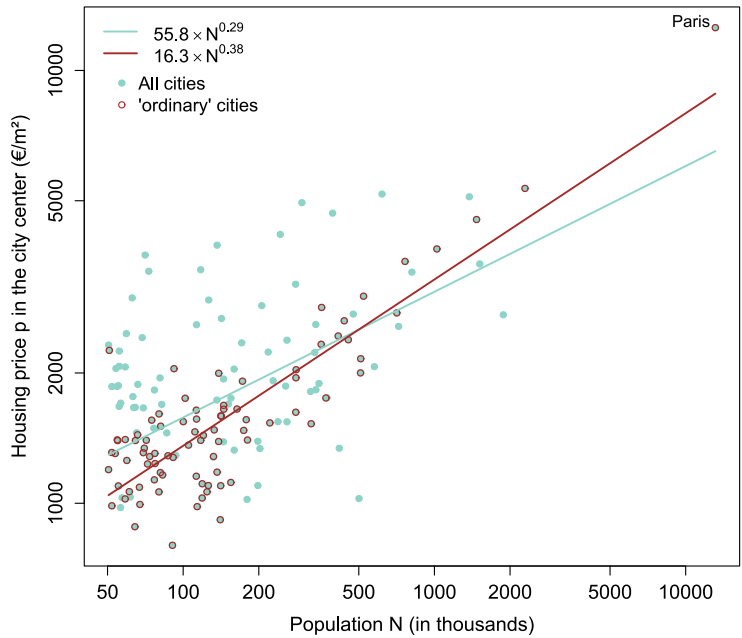

*Figure S7. Median housing price in the center ($0 < r < 1\ km$) of French cities as a function of city population. Regressions (no weighting) are displayed for all cities and 'ordinary' cities only.*

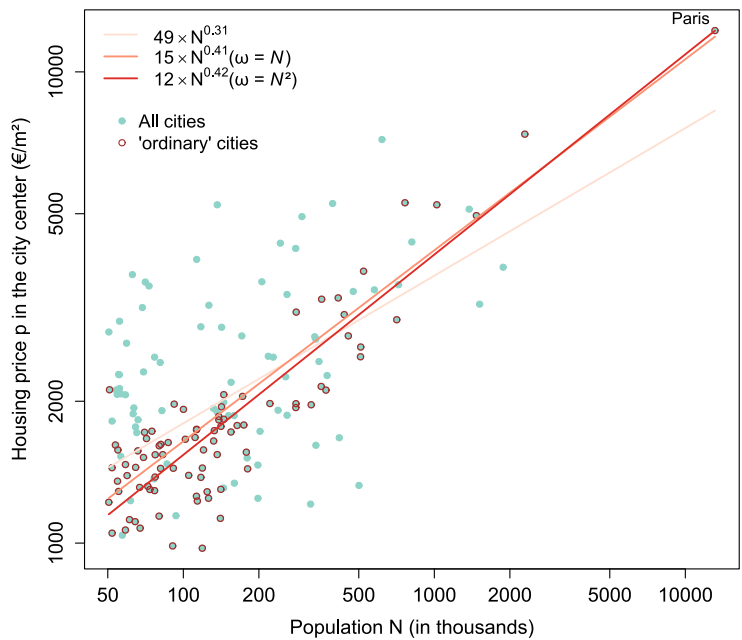

*Figure S8. Median housing price in the center ($0 < r < 1\ km$) of French cities (**houses only**) as a function of city population.*

## 6. Fluctuations

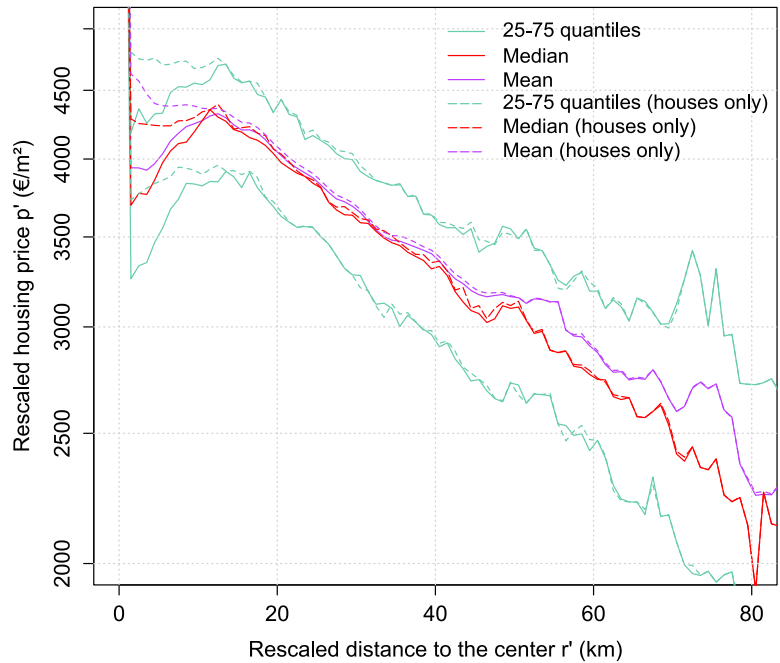

*Figure S9. Distributions of housing price rescaled radial profiles for 'ordinary' cities only. Fluctuations are displayed for both apartments and houses (solid lines) and for houses only (dashed lines)*

## 7. Optimal scaling exponents

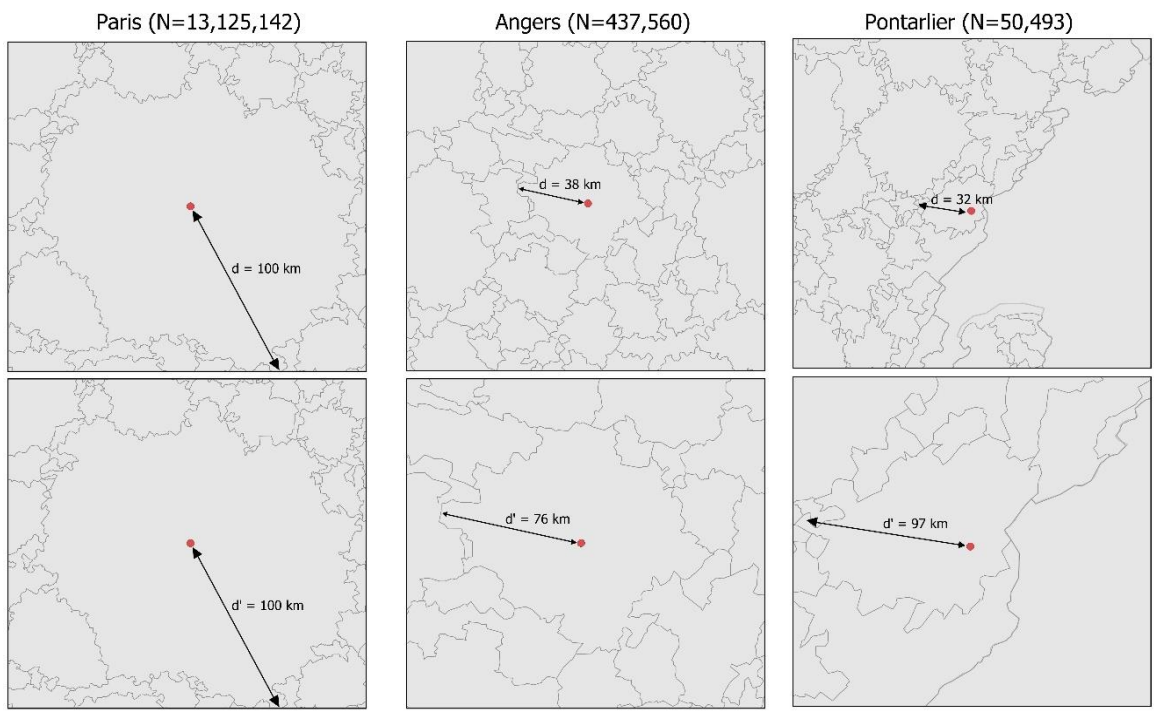

*Figure S10. Maximum distance from the center to the periphery before (first row) and after (second row) rescaling for cities of different sizes.*

## 8. Non-radial urban structure

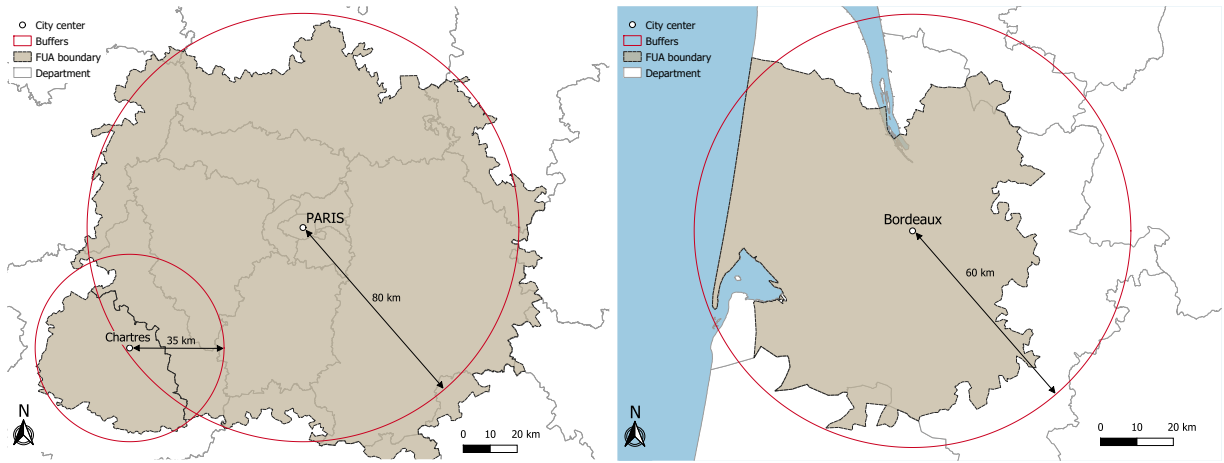

*Figure S11. Map of the urban areas of Chartres and Paris, and Bordeaux*

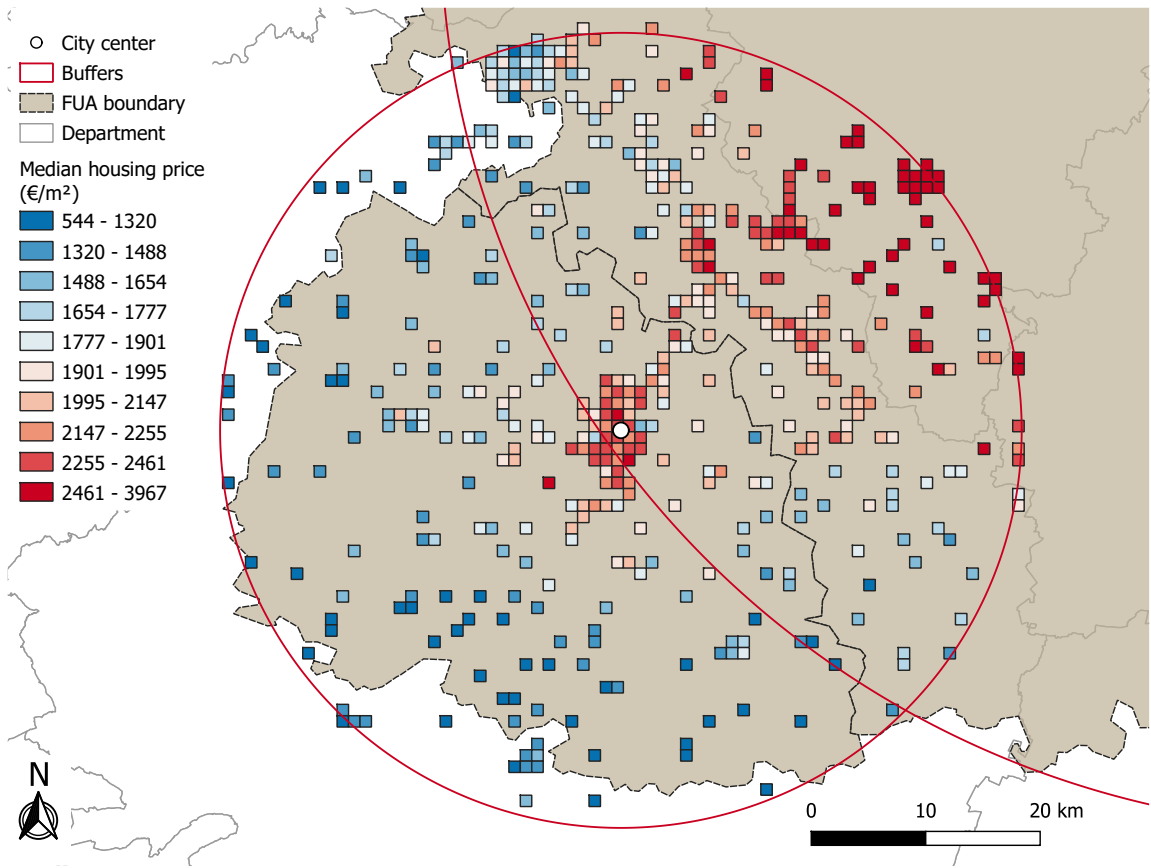

*Figure S12. Housing prices within the urban areas of Chartres*

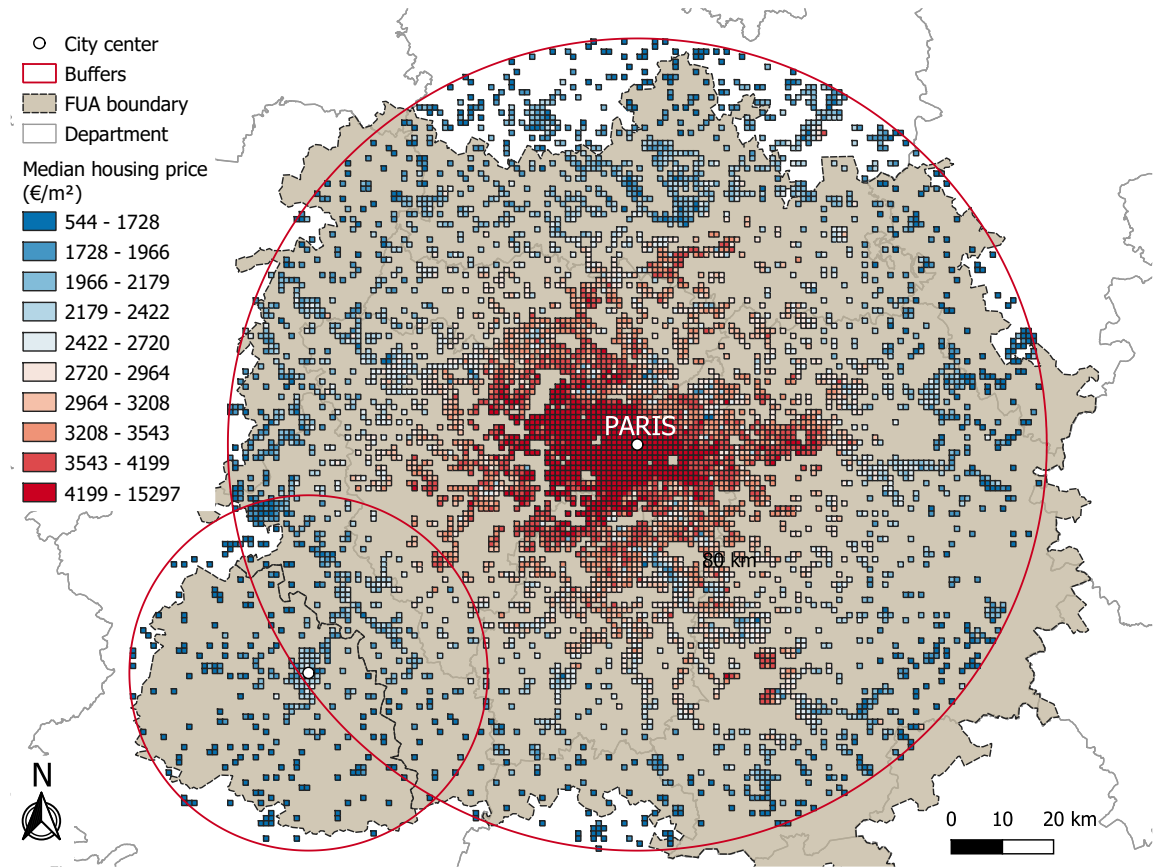

*Figure S13. Housing prices within the urban areas of Paris and Chartres*

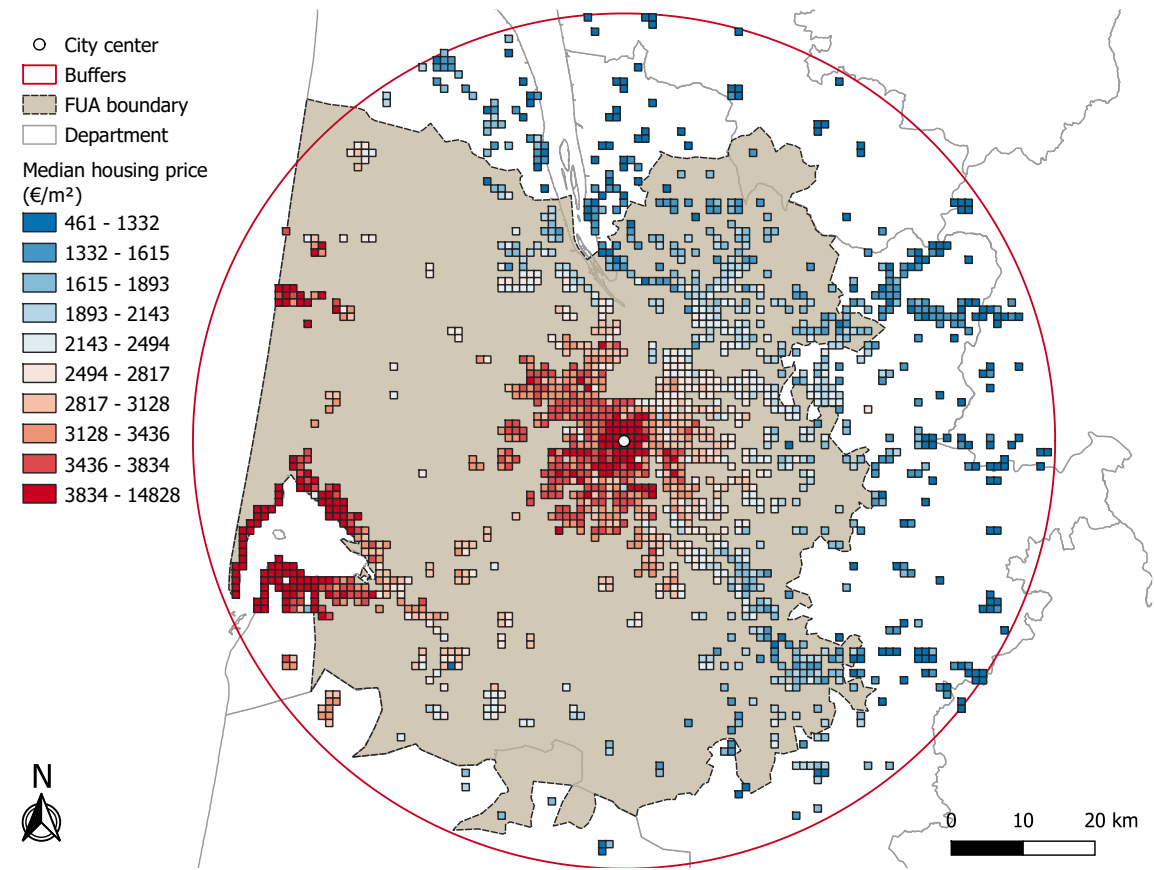

*Figure S14. Housing prices within the urban areas of Bordeaux*

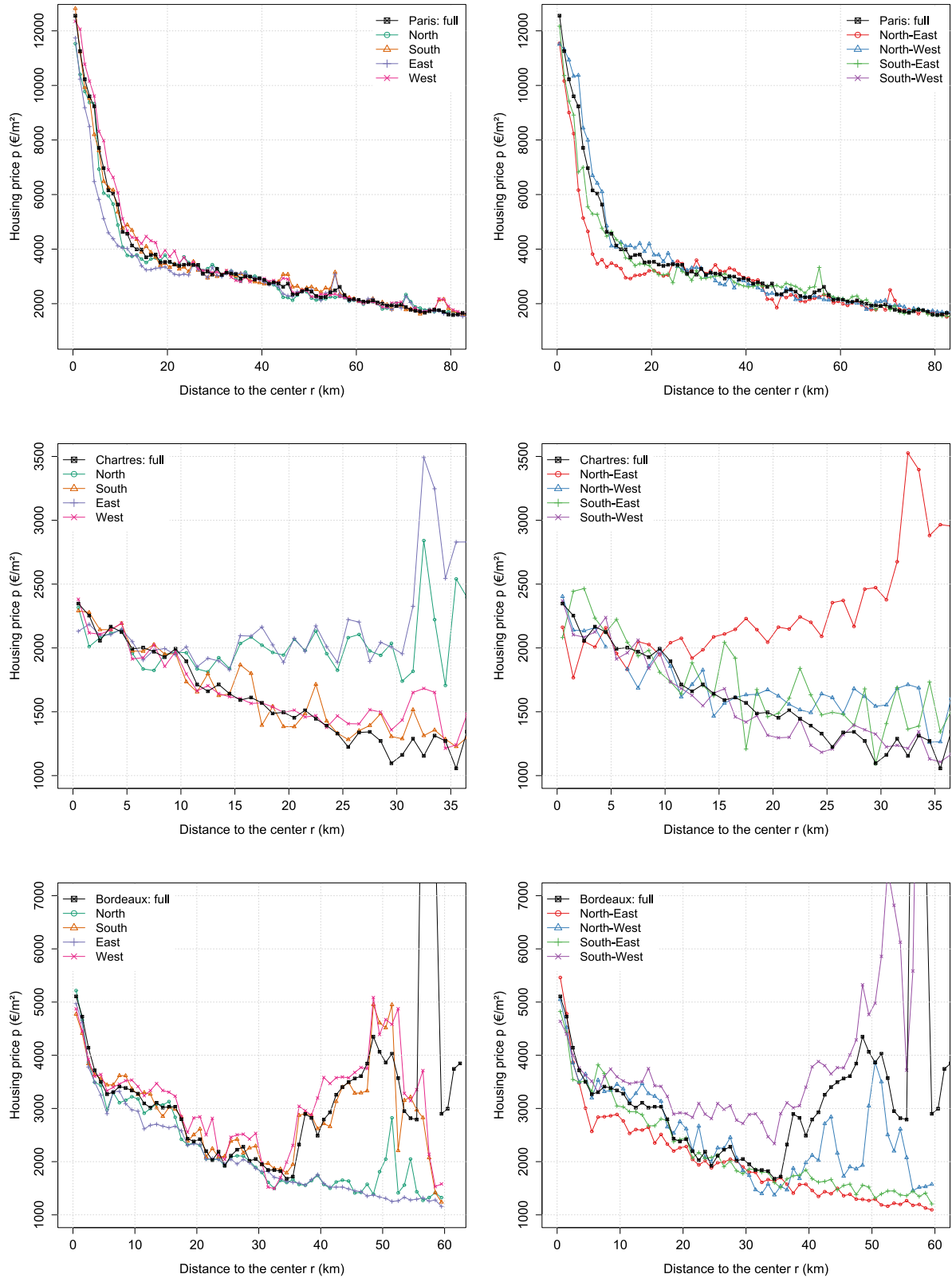

*Figure S15. Angular variations of housing prices radial profiles for Paris (first row), Chartres (second row) and Bordeaux (third row). Panels show radial profiles with half-discs oriented East, West, North and South (left column), and with quarter-discs oriented North-East, South-East, North-West and South-West (right column).*